\documentclass[prb,twocolumn,showpacs,superscriptaddress]{revtex4-1}
\usepackage{epsf}
\usepackage{psfig}
\usepackage{xfrac}
\usepackage{graphicx}
\usepackage{color}
\begin{document}

\title {Large ground state magnetic moment and magnetocaloric effect in Ni$_{2}$Mn$_{1.4}$In$_{0.6}$}

\author{Sanjay Singh}
\email[]{Sanjay.Singh@cpfs.mpg.de}
\affiliation{Max Planck Institute for Chemical Physics of Solids, N\"{o}thnitzer Stra{\ss}e 40, D-01187 Dresden, Germany}
\author{Luana Caron}
\email[]{Luana.Caron@cpfs.mpg.de}
\affiliation{Max Planck Institute for Chemical Physics of Solids, N\"{o}thnitzer Stra{\ss}e 40, D-01187 Dresden, Germany}
\author{S. W. D'Souza}
\affiliation{Max Planck Institute for Chemical Physics of Solids, N\"{o}thnitzer Stra{\ss}e 40, D-01187 Dresden, Germany}
\author{Tina Fichtner}
\affiliation{Max Planck Institute for Chemical Physics of Solids, N\"{o}thnitzer Stra{\ss}e 40, D-01187 Dresden, Germany}
\author{Giacomo Porcari}
\affiliation{Department of Physics and Earth Sciences, Parma University, Viale G.P. Usberti n.7/A (Parco Area delle Scienze) 43124 - Parma, Italy}
\author{C. Shekhar}
\affiliation{Max Planck Institute for Chemical Physics of Solids, N\"{o}thnitzer Stra{\ss}e 40, D-01187 Dresden, Germany}
\author{S. Chadov}
\affiliation{Max Planck Institute for Chemical Physics of Solids, N\"{o}thnitzer Stra{\ss}e 40, D-01187 Dresden, Germany}
\author{Massimo Solzi}
\affiliation{Department of Physics and Earth Sciences, Parma University, Viale G.P. Usberti n.7/A (Parco Area delle Scienze) 43124 - Parma, Italy}
\author{Claudia Felser}  
\affiliation{Max Planck Institute for Chemical Physics of Solids, N\"{o}thnitzer Stra{\ss}e 40, D-01187 Dresden, Germany}

\begin{abstract}
A large conventional magnetocaloric effect at the second order magnetic transition  in cubic  Ni$_{2}$Mn$_{1.4}$In$_{0.6}$  Heusler alloy is reported. The isothermal magnetization at 2\,K shows a huge ground state magnetic moment of about 6.17 $\mu$B$/f.u$. The theoretical calculations show that the origin of the large magnetic moment in cubic Ni$_{2}$Mn$_{1.4}$In$_{0.6}$ results from the strong ferromagnetic interaction between Mn- Ni and Mn-Mn sublattices. The experimental magnetic moment is in excellent agreement with the moment calculated from the theory.  The large magnetic moment gives rise to considerably high adiabatic temperature and entropy changes at the magnetic transition.  The present study opens up the possibility to explore cubic Heusler alloys for magnetocaloric applications.

\end{abstract}


\maketitle

Magnetic materials showing large magnetocaloric effect (MCE) have recently gained vast interest due to their potential use in magnetic refrigeration technology. The MCE is intrinsic to all magnetic materials, being particularly high around second and first order magnetic phase transitions.\cite{Pecharsky97, Annaorazov92, Wada01, Tegus02, Dung11, HuPRB01, KrenkeNM05, Liu12, Pareti03}  Since the observation of the so-called giant MCE around room temperature in Gd$_{5}$Ge$_{2}$Si$_{2}$\cite{Pecharsky97}, materials showing first-order magnetic phase transitions have been investigated intensively aiming at cooling applications. However, the practical application of giant MCE materials is necessarily hindered by the nature of the phase transition itself. In order to drive a first order phase transition energy must be spent to overcome the energy barrier between states. This leads to intrinsic irreversibilities in both entropy and adiabatic temperature changes which can drastically reduce cooling efficiency. Moreover, in the case of strong first order phase transitions where the crystal lattice discontinuity is the result of a symmetry change, if often leads to materials that are physically unstable upon thermomagnetic cycling.\cite{Cesari}

In contrast, materials presenting second order magnetic phase transitions show lower entropy changes due to the continuous nature of the transition. However, a second order continuous phase change is fully reversible. For example, Gd is used as the benchmark material in magnetic refrigerator prototypes.\cite{Brown76} Gd is a rare earth metal which shows a second order ferro to paramagnetic phase transition at (~294\,K) and derives its considerably high entropy change from the high 4d moments ( M$_{S}$ ~7.5\,$\mu$B$/f.u$). However, the high costs of Gd make its use in commercial applications prohibitive.\cite{Yu10}

Ni-Mn based magnetocaloric Heusler alloys have emerged as important candidates for cooling applications due to their low cost and environmental friendliness compared to rare-earth based materials.\cite{HuPRB01, KrenkeNM05, Liu12, Pareti03, SSingh14a, SSingh14} In these alloys the strong coupling between magnetic and structural degrees of freedom at the martensitic phase transition generates large entropy and adiabatic temperature changes. However, the martensitic phase transition is crossed at a high energy cost which translates into large hysteresis losses and low efficiency for cooling applications. It is necessary to drastically reduce thermal hysteresis to make real use of these materials.  Here we report the large conventional magnetic entropy change near room temperature at the second order magnetic transition in the Ni$_{2}$Mn$_{1.4}$In$_{0.6}$  Heusler alloy, due to its large ground state magnetic moment of about 6.17 $\mu$B$/f.u$. The ground state magnetic moment observed for this compound is comparable to that of Gd and the highest ground state moment observed in the magnetocaloric Heusler family. 

A polycrystalline ingot of Ni$_{2}$Mn$_{1.4}$In$_{0.6}$ was prepared by melting appropriate quantities of Ni, Mn and In of purity higher than 99.99\% in an arc furnace. The resulting ingot was then annealed at 1173\,K for 21 days for homogenisation and subsequently quenched into ice water. The room-temperature crystal structure was obtained by powder X-ray diffraction (XRD) using CuK$\alpha$ radiation. Prior to the XRD data collection, the powder obtained from grinding the ingots  was further annealed as to remove residual strain.\cite{Ranjan09,Singh10,Singh14} XRD data analysis was performed using the Rietveld refinement method as implemented in the JANA2006 software package.\cite{Jana} 
Sample composition was determined by energy dispersive X-ray spectroscopy (EDX). 
The sample was found to be homogeneous  with an average composition of Ni$_{1.97}$Mn$_{1.4}$In$_{0.63}$, which will be refered as Ni$_{2}$Mn$_{1.4}$In$_{0.6}$ henceforth. Isofield and isothermal magnetisation measurements were performed in a Quantum design SQUID-VSM magnetometer. 

The reversible adiabatic temperature change measurements were carried out using a Cernox bare chip thermoresistor as temperature sensor. The experiment was performed in a Optistat CF-V cryostat by Oxford Instruments. The sample (30\,mg) was glued on the top of the "cold finger" (a layer of Kapton was placed in between to slow down the heat transfer between sample and cryostat) while the temperature sensor was glued on the top surface of the sample. The setup is inserted between the poles of a low inductive electromagnet (max field 2\,T, magnetic field rise time 1\,s, see Porcari et al.\cite{Porcari}) The sample's temperature is acquired while the cryostat sweeps across the transition region at a rate of 1\,K/min and the magnetic field is continuously switched on and off.

The self-consistent band structure calculations were carried by spin polarized non-relativistic Korringa-Kohn-Rostocker (SPRKKR) Green's function technique \cite{Ebert11}. The exchange and correlation effects were incorporated within the generalized gradient approximation (GGA) in the Perdew-Burke-Ernzerhof parametrization scheme. \cite{Perdew96} Brillouin zone integrations were performed on a 22$\times$22$\times$22 mesh of $k$-points. The angular momentum expansion up to l$_{\rm max}$=~3 has been used for each atom.  The Substitutional disorder in Ni$_2$Mn$_{1.4}$In$_{0.6}$, has been  taken into account by employing the coherent potential approximation (CPA) method as implemented in the SPRKKR code.  Both the energy convergence criterion and the CPA tolerance were set to 10$^{-5}$ Ry.


\begin{figure}[htb]
\includegraphics[trim = 25mm 0mm 0mm 5mm, clip, scale=0.5]{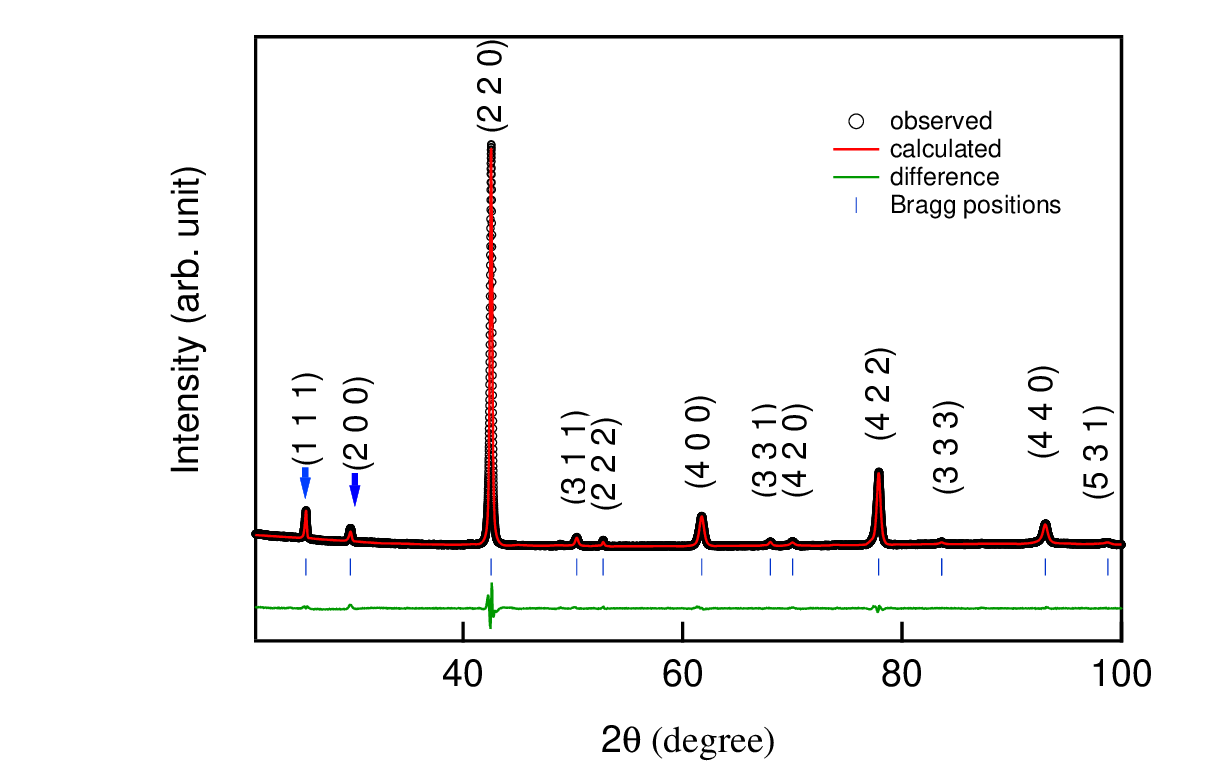}
\vspace{-18pt}
\caption{Room-temperature XRD pattern of  Ni$_{2}$Mn$_{1.4}$In$_{0.6}$. Observed, calculated and residue patterns are shown by black dots, red solid line and green solid line, respectively. Blue ticks represents Bragg peak positions. Blue arrows indicate the superstructure reflections of the L2$_1$ order phase.} 
\label{XRD}
\end{figure} 



\begin{figure}[htb]
\includegraphics[scale=0.36]{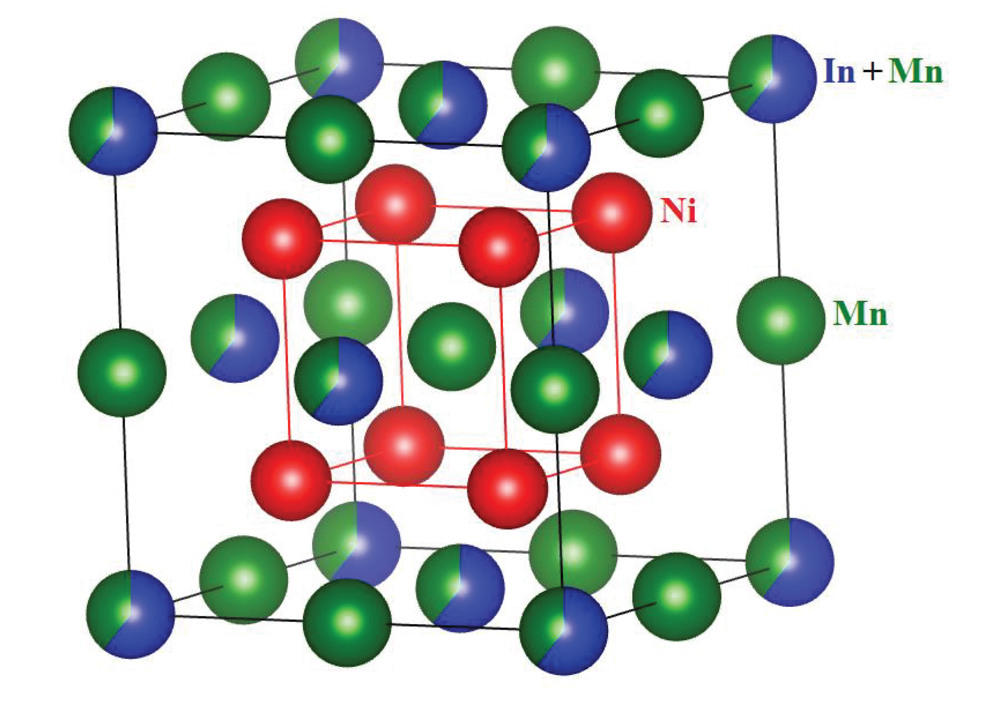}
\vspace{-8pt}
\caption{Crystal structure of the cubic Ni$_{2}$Mn$_{1.4}$In$_{0.6}$ Heusler alloy.} 
\label{austenite_cell}
\end{figure}

The Rietveld refinement of the room-temperature XRD pattern is shown in Fig.\ref{XRD}. The Rietveld refinement was performed using the L2$_1$ structure (space group Fm$\overline{3}$m) where Ni atoms  occupy the (0.25,0.25,0.25) and (0.75,0.75,0.75) positions, Mn the (0.5, 0.5, 0.5)  and In the (0 0 0) positions, see Fig.\ref{austenite_cell}. Excess Mn atoms occupy the In position. The sample is found to be single phase as all observed Bragg reflections could be indexed using the cubic L2$_1$. The refined lattice parameter was 6.004 \AA. The observation of superstructure Bragg reflections (111) and (200) (Blue arrows in Fig.\ref{XRD}) correspond to the L2$_1$  ordering .\cite{SSingh14}

\begin{figure}[htb]
\includegraphics[scale=0.86]{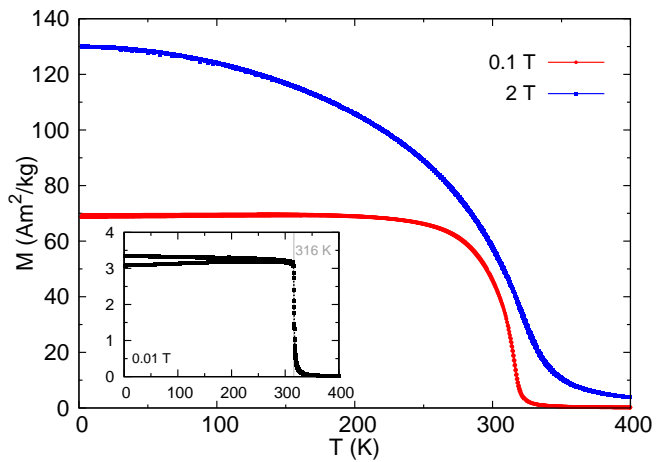}
\vspace{-15pt}
\caption{Zero-field-cooled (ZFC), field-cooled-cooling (FCC) and field-cooled-warming (FCW) magnetization curves at different applied magnetic fields. Inset shows the magnetization at low magnetic field of 0.01\,T and T$_C$ is indicated.} 
\label{MxT}
\vspace{-15pt}
\end{figure}

\begin{figure}[h]
\includegraphics[scale=0.86]{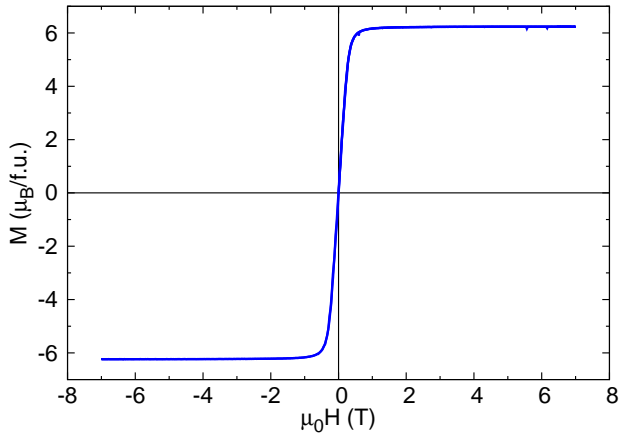}
\vspace{-18pt}
\caption{The isothermal magnetization curve at 2 K showing saturation moment.} 
\vspace{-3pt}
\label{MxH@2K}
\vspace{-10pt}
\end{figure}

\begin{figure}[htb]
\vspace{-10pt}
\includegraphics[scale=0.86]{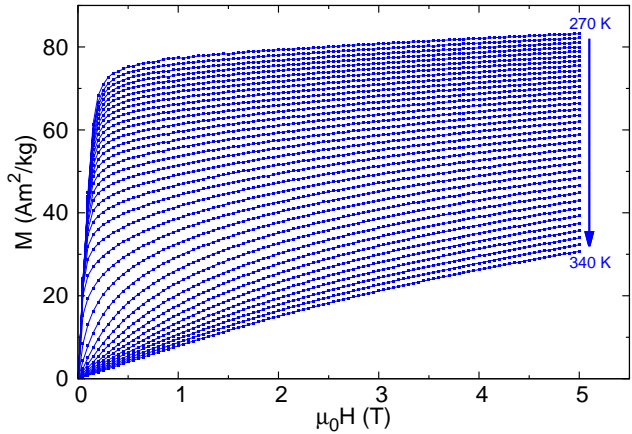}
\vspace{-18pt}
\caption{The isothermal M(H) curves within the temperature range of 270\,- 340K at 2\,K temperature.} 
\label{MxHisoT}
\vspace{-5pt}
\end{figure}

\begin{table*} [t]
\caption{Starting and converged Mn, Ni spin magnetic moments ($\mu_{B}$/atom) and the corresponding converged total energies (E$_{tot}$, meV) of Ni$_2$Mn$_{1.4}$In$_{0.6}$.  The lowest E$_{tot}$ is taken to be 0 meV as a reference.}
\centering
~\\
\begin{tabular}{l c c c c c c} \hline\hline
   & \multicolumn{2} {c } {Starting Mn, Ni moment} & \multicolumn{4} {c} {Converged Mn, Ni moment}  \\ \hline\hline
      & {ferromagnetic} & {antiferromagnetic} & {ferromagnetic} &  &     &{antiferromagnetic}   \\ \hline
 Mn$_{\rm In}$ & 3.00 & -3.00 & 3.91 &    &     &-4.07 \\  
 Mn$_{\rm Mn}$ & 3.00 &  3.00 & 3.75 & &       &3.76  \\  
      Ni       & 0.3  &  0.3  & 0.49 & &           &0.26  \\  \hline
    &  & Total converged energy (E$_{tot}$) &  0  & &        &252   \\ \hline\hline
 \end{tabular}
\label{1mn1.4_lowest_energy}
\end{table*} 
The magnetization curves as a function of temperature for different applied magnetic fields are shown in Fig.\ref{MxT}.  A sharp change in magnetization is observed at ~316\,K, which corresponds to the magnetic order-disorder transition temperature (T$_C$). This transition is expected to be second order in nature since no hysteresis is observed even at low fields (see inset of fig. \ref{MxT}) and the magnetisation increases monotonically with increasing field under isothermal conditions (see Fig. \ref{MxHisoT}). 
Notice that, no transition to a lower magnetic moment is observed below T$_C$, as would be expected from reports for a similar composition.\cite{Moya07} This indicates that the present sample does not undergo a martensitic phase transition and retains its cubic structure down to 2\,K. Furthermore the absence of considerable difference between ZFC and FC magnetization curves at low field (0.01 T) clearly indicates that the antiferromagnetic interactions usually observed in these alloys are not present in this sample, which further indicates the absence of any Ni-Mn antisite disorder.\cite{SSingh12}

Fig.\ref{MxH@2K} shows the isothermal magnetization measured at 2\,K. It is interesting to note that the sample has a large ground state saturation moment of about 6.17 $\mu$B$/f.u$, which indicates strong ferromagnetic interactions in this alloy. Recently a large saturation moment of ~6 $\mu$B$/f.u$ has been reported in Ni-Mn based shape memory alloys under pulsed-high magnetic fields up to 60\, T at 4.2\,K.\cite{Nayak14} It has been proposed that the large magnetic moment is related to the field-induced martensite to austenite transition at higher magnetic fields. 
The present observation clearly shows that the 6.17 $\mu$B$/f.u$ magnetic moment is due to the austenite phase. 

SPRKKR calculations for Ni$_2$Mn$_{1.4}$In$_{0.6}$ were performed in the cubic L2$_1$ structure (see Fig.\ref{austenite_cell}) using the experimental lattice parameter and atomic positions as determined from XRD. The Ni atoms occupied on the 8c [(0.25, 0.25, 0.25) and (0.75, 0.75, 0.75)] sites. Mn atoms are at the 4b (0.5, 0.5. 0.5) site. The 40$\%$ excess Mn atoms in Ni$_2$Mn$_{1.4}$In$_{0.6}$ are occupied at the In 4a (0, 0, 0) site. The Mn atoms at the 4b site are referred to as Mn$_{\rm Mn}$ {\it i.e.} Mn atoms in Mn site. In these systems, the self-consistent calculations might not converge to the actual magnetic ground state,\cite{Barman08,Chakrabarti09,SSingh12}  therefore, In order to obtain the accurate lowest energy magnetic state we have performed total energy (E$_{tot}$) calculations starting with ferromagnetic and antiferromagnetic collinear spin configurations of the Mn atoms at the 4a and 4b sites.
The converged spin moments of  Mn$_{\rm In}$, Mn$_{\rm Mn}$ and Ni atoms in both spin configurations clearly indicates that the moments are localised on the Mn atoms (Table~\ref{1mn1.4_lowest_energy}), as expected for Heusler alloys.  The initial {\it ferromagnetic}  Mn starting spin configuration of 3~$\mu_B$ shown in the Table~\ref{1mn1.4_lowest_energy} corresponds to the lowest E$_{tot}$  with a total spin moment of  6.26 $\mu_{B}$/f.u. On the other hand, for the {\it antiferromagnetic} starting spin configuration, the calculations converge to a ferrimagnetic ground state (see Table~\ref{1mn1.4_lowest_energy}) resulting in 252 meV larger total energy with the Mn$_{\rm Mn}$ and Mn$_{\rm In}$ moments of 3.76 and -4.07 $\mu_{B}$, respectively leading to a total moment of 2.62 $\mu_{B}$/f.u. Thus, the SPRKKR calculations are in excellent agreement with the magnetisation measurements (M = 6.17 $\mu_{B}$/f.u., Fig.\ref{MxH@2K}) indicating a ferromagnetic ground state.

\begin{figure} [htb]
\begin{center}
\includegraphics[trim = 10mm 5mm 10mm 5mm, clip, scale=0.46]{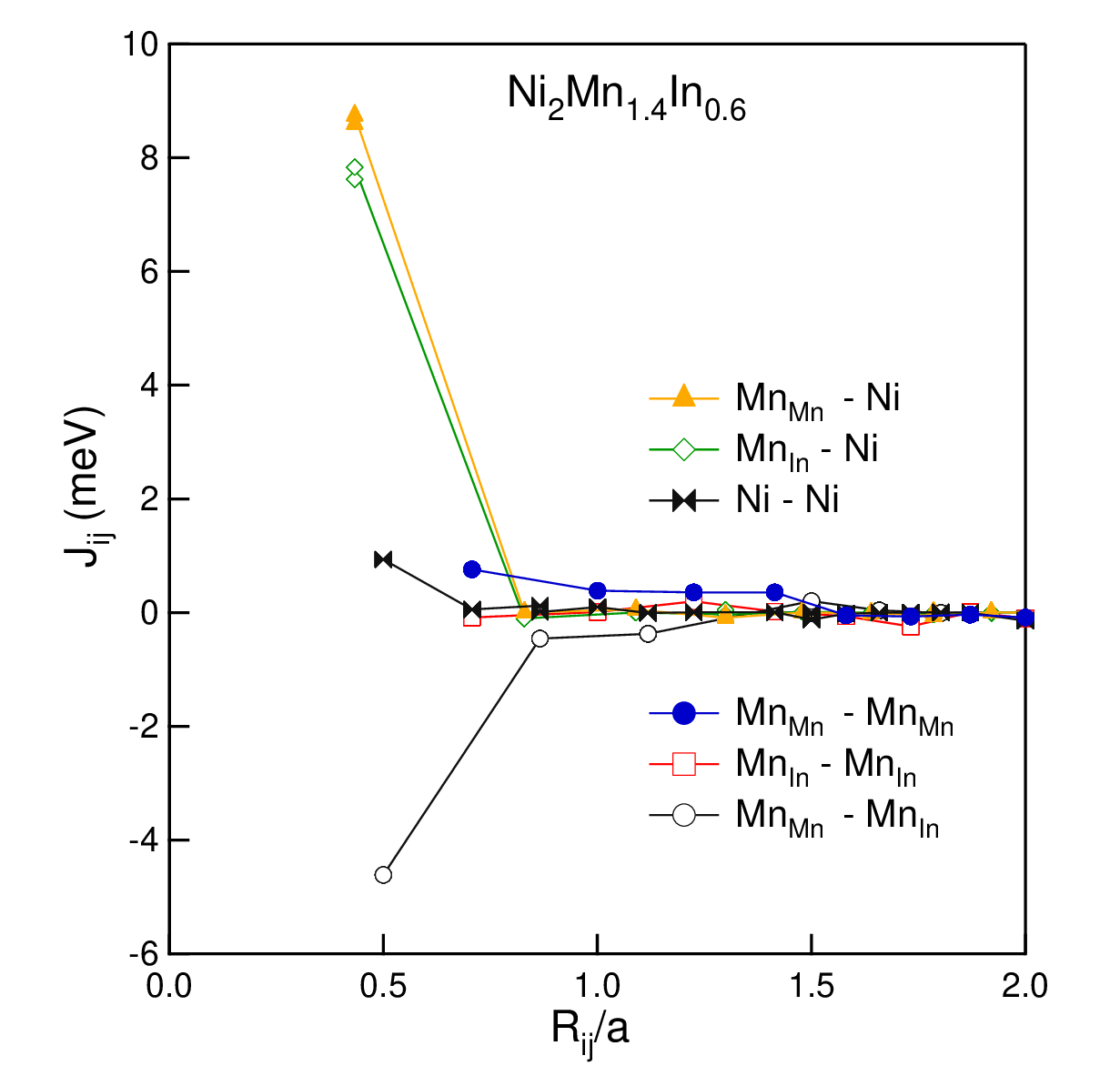}
\end{center}
\vspace{-20pt}
\caption{(color online) Magnetic exchange interaction parameters $J_{ij}$ of Ni$_2$Mn$_{1.4}$In$_{0.6}$ between an atom $i$ and its neighbor $j$ located at a distance of R$_{ij}$. The distances are normalized with respect to the lattice parameter $a$.}
\label{2exchange}
\vspace{-10pt}
\end{figure}

The magnetic exchange coupling parameters (J$_{ij}$) of Ni$_2$Mn$_{1.4}$In$_{0.6}$ calculated based on the Green's function method by using the formulation of Liechtenstein {\it et al.}\cite{Liechtenstein84} are plotted in Fig.~\ref{2exchange}. The $J_{ij}$ calculations are performed within a cluster of radius 3$a$, where $a$ is the lattice parameter. A coexistence of antiferromagnetic (Mn$_{\rm Mn}$- Mn$_{\rm In}$, Mn$_{\rm In}$- Mn$_{\rm In}$, ) and ferromagnetic (Mn$_{\rm Mn}$- Ni, Mn$_{\rm In}$- Ni, Ni-Ni and Mn$_{\rm Mn}$- Mn$_{\rm Mn}$) interactions between the nearest neighbors (nn) in Ni$_2$Mn$_{1.4}$In$_{0.6}$ is evident form the Fig.~\ref{2exchange}. In the first nn the dominant antiferromagnetic interaction is displayed by the Mn$_{\rm Mn}$- Mn$_{\rm In}$ (-4.6 meV) sublattice with a subsequent decrease in the antiferomagnetic strength extending upto third nn (-0.37 meV) . The antiferromagnetic Mn$_{\rm In}$- Mn$_{\rm In}$ (-0.1 meV) interaction which is found to be significantly weaker in the first nn exhibits a crossover to the ferromagntic interaction in its third coordination shell resulting in a magnetic exchange energy of about 0.2 meV. This exhibits a long range oscillatory behavior associated with the Mn$_{\rm In}$- Mn$_{\rm In}$ sublattice. A comparison of all sublattice exchange interactions at the first nn  displayed in Fig.~\ref{2exchange} clearly shows that the dominant interaction in Ni$_2$Mn$_{1.4}$In$_{0.6}$ is ferromagnetic in nature. The dominating contribution to the ferromagnetic interaction arises from the Mn$_{\rm Mn}$- Ni sublattice interaction, which amounts to 8.7 meV in the first nn.  The  exchange interaction energy for the Mn$_{\rm In}$- Ni sublattice corresponds to 7.8 meV in the first nn and vanishes for more distant neighbors. The first nn interactions of the Ni-Ni (0.9 meV) and Mn$_{\rm Mn}$- Mn$_{\rm Mn}$ (0.8 meV) sublattices are significantly smaller in comparison with the Mn$_{\rm Mn}$- Ni interaction.  In contrast, the Mn$_{\rm Mn}$- Mn$_{\rm Mn}$ sublattice interaction is found to possess considerable ferromagnetic strength up to the fourth coordination shell (0.36 meV) resulting in a long range ferromagnetic behavior. This indicates that the Mn$_{\rm Mn}$- Mn$_{\rm Mn}$ interaction along with Mn$_{\rm Mn}$- Ni interaction plays a significant role in stabilizing the ferromagnetism in Ni$_2$Mn$_{1.4}$In$_{0.6}$.

The sharp magnetization change at T$_C$ and extremely high ground state magnetic moment are good indications for a high MCE. Therefore we studied the magnetocaloric behavior of this alloy.

\begin{figure}[t]
\includegraphics[scale=0.86]{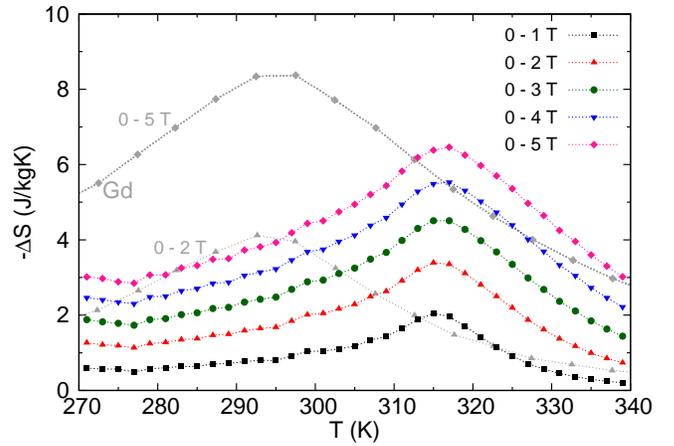}
\vspace{-20pt}
\caption{Magnetic entropy change  $\Delta\, S$ as a function of temperature under different applied fields. The $\Delta\, S$ of Ni$_{2}$Mn$_{1.4}$In$_{0.6}$  is compared with Gd (after Tegus et al.\cite{Tegus02}). }
\label{dSxT}
\vspace{-10pt}
\end{figure}

The isothermal magnetization curves $M(H)_T$ for Ni$_{2}$Mn$_{1.4}$In$_{0.6}$  in a temperature range between 270\,K and 340\,K measured every 2\,K  in temperature and field increase mode are shown in Fig.\ref{MxHisoT}. The change in magnetic entropy ($\Delta S$) is calculated from the measured $M(H)_T$ curves using the Maxwell relations.\cite{Pecharsky97} 

The entropy change curves are broad and show maxima at 3.3 \,J/kgK and 6.3\,J/kgK for 0 - 2 T and 0 - 5 T field changes, respectively (see Fig. \ref{dSxT}). These values are comparable to those observed for Gd\cite{Tegus02}   around the same temperature range, with the advantage that this compound is rare earth-free. 

In order to fully assess the material's properties for applications direct adiabatic temperature change measurements were also performed. We observed adiabatic temperature changes around 1.5 and 2\,K for 0 - 1\,T and 0 - 2\,T magnetic field changes, respectively.



\begin{figure}
\includegraphics[scale=0.86]{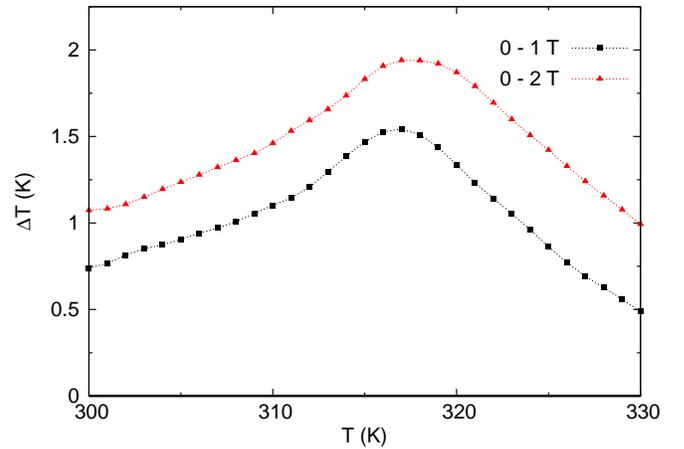}
\vspace{-18pt}
\caption{Adiabatic temperature change as a function of temperature at selected values of the magnetic field.} 
\label{dTxT}
\vspace{-18pt}
\end{figure} 
It should be noted that the entropy and adiabatic temperature changes obtained for this sample are consistently higher than those obtained for a very similar composition by Moya et al.\cite{Moya07}. This may be  due to the presence of antisite disorder, which gives origin to the Mn-Mn antiferromagnetic interactions and results in the maximum magnetic moment observed by Moya and co-workers being drastically lower than what we observe, reaching a maximum of 75\,Am$^2$/kg for an applied field of 1\,T. In fact, the saturation magnetization of 6.17\,$\mu$B$/f.u$ observed for this sample matches remarkably well the ab initio predicted value of 6.26\,$\mu$B$/f.u$. Although the adiabatic temperature change values obtained for the present alloy are lower than those of Gd metal,\cite{dankov} they are comparable to those observed for some shape memory Heusler alloys at their first order martensitic transition\cite{khovailoprb, pasqule,Acet}. However low, the adiabatic temperature change observed around a second order magnetic phase transition is fully reversible. While high values have been reported for several Heusler alloys around the first order martensitic phase transition, the effect is mostly irreversible due to thermal hysteresis making these materials inappropriate for applications.\cite{Liu12}

In summary we report a large conventional magnetocaloric effect at the second order magnetic phase transition in a well ordered cubic Heusler alloy. We observed a huge saturation magnetization of 6.17\,$\mu$B$/f.u$, which is the highest ever obtained in the Ni-Mn based magnetocaloric Heusler system to the best of our knowledge. Our calculations explain the observed high magnetic moment and point the way on how to obtain large saturation moments and consequently high MCE in Ni-Mn-based Heusler compounds.  The zero hysteresis and full reversibility are major advantages of second order phase transitions over the first order processes. In Heusler alloys, the magnetic moment as well as the magnetic transition temperature can be easily tuned by varying composition. Therefore the present results open up the possibility to investigate cubic Heusler alloys for rare earth free magnetic-cooling applications.


This work was financially supported by the Deutsche Forschungsgemeinschaft DFG (Project No. TP 2.3-A of Research Unit FOR 1464ASPIMATT) and by the ERC Advanced Grant (No. 291472) "Idea Heusler". S. S. thanks Alexander von Humboldt for fellowship.


\end{document}